\documentstyle[preprint,aps]{revtex}
\begin{document}
\title{Planar Superconductivity via Kosterlitz-Thouless mechanism}
\author{R. MacKenzie}
\address{ Laboratoire de Physique Nucl\'{e}aire, \\
Universit\'{e} de Montr\'{e}al, Montr\'{e}al, Qu\'{e} H3C 3J7, {\bf CANADA}, \\
{\rm P.K. Panigrahi} \\
School of Physics, University of Hyderabad\\
Hyderabad - 500 046, {\bf INDIA} \\   and \\
{\rm S.Sakhi} \\
Department of Physics, University of British Colombia, \\
Vancouver, B.C. V6T2A6, {\bf CANADA} \\ }
\maketitle
\begin{abstract}
The phase structure of a $(2+1)$ - dimensional model of
relativistic fermions with a four fermi interaction is analyzed in the
strong coupling regime using the large $N$ perturbation theory. It is shown
that, this model exhibits a low temperature superconducting phase due to the
vortex-anti vortex binding via Kosterlitz-Thouless mechanism. Above a
critical temperature, vortices unbind and superconductivity is destroyed;
at a still higher temperature the vacuum expectation value of a neutral
order parameter vanishes. The ground state respects parity and time
reversal symmetries. \\
\end{abstract}
\newpage

\noindent{\bf 1.~~ Introduction} \\
Superconductivity in planar field theoretical models  have aroused
considerable interest in recent times because of their relevance to
condensed matter systems.$\rm^1$ A thorough analyses of the phase
structure of these models is necessitated due to the fact that
the conventional explanation based on the spontaneous breaking of a
$U(1)$ symmetry may not be applicable to the lower dimensional world.
In $2+1$ dimensions the vacuum expectation value (VEV) of a charged order
parameter vanishes for non-zero temperatures due to infrared divergences,
much akin to the celebrated Coleman-Mermin-Wagner$\rm^2$ theorem, which
states that spontaneous breaking of a continuous symmetry does not take
place in $1+1$ dimensions. In light of the above constraint, one usually
invokes a small interplanar coupling or takes recourse in
unconventional mechanisms e.g. parity and time reversal violating anyon
superconductivity.$\rm^3$
This note briefly describes an alternate model$\rm^4$, involving relativistic
fermions with a four-fermi interaction, which exhibits
superconductivity due to the vortex confinement mechanism of
Kosterlitz and Thouless(KT).$\rm^5$ Relativistic fermions appear as the
relevant low energy degrees of freedom in the Hubbard and related models,
$\rm^6$ widely believed to be of relevance to high $T_c$ superconductors and
four-fermi couplings also arise naturally in these models in the presence of
doping.$\rm^7$ Without a precise knowledge about these couplings, we
consider a four-fermi term of BCS type and study the various phases, taking
advantage of the model's large $N$ renormalizability.$\rm^8$

The paper is organized as follows. Section 2 is devoted to the
computation of the low energy effective action in parallel to the
derivation of the  Landau-Ginzburg effective action in the BCS theory.
It is shown that a neutral order parameter can have a nonvanishing
VEV up to a critical temperature $T_0$, although the VEV of a charged
order parameter vanishes for $T\neq0$. In Sec.3, we study the occurrence
of superconductivity in this model, after taking in to account the dynamics
of the phase degrees of freedom. We conclude in Sec.4 with some comments. \\

\newpage
\noindent
{\bf 2.~~ Low Energy Effective Action} \\
Calculation of the low energy effective action which encapsulates the relevant
dynamical degrees of freedom, has been a common approach since the early days
of superconductivity. Physically, there is a gap in the fermion spectrum
below the  critical temperature, because of which long wavelength excitations
of the condensate are energetically cheaper to create than fermionic
excitations. The effective action exhibits the spontaneous breaking
of the $U(1)$ symmetry responsible for superconductivity. Below, we will
outline the derivation of the effective action for the BCS
Lagrangian$^{\rm 9}$,

\begin{equation}
{\cal  L_{BCS}} =  \psi^\dagger_\uparrow  (i  \partial_t  -  \epsilon (p))
\psi_\uparrow   +  \psi^\dagger_\downarrow  (i  \partial_t - \epsilon (p))
\psi_\downarrow   - \lambda \psi^\dagger_\uparrow \psi^\dagger_\downarrow
\psi_\downarrow \psi_\uparrow \, \, \, \, ,
\end{equation}

\noindent
and follow the same procedure for the model presented here. To get
the effective theory, one introduces auxiliary fields to rewrite the ${\cal
L}_{BCS}$ as,

\begin{equation}
{\cal  L}_1 =  \psi^\dagger_\uparrow  (i  \partial_t  -  \epsilon (p))
\psi_\uparrow   +  \psi^\dagger_\downarrow  (i  \partial_t - \epsilon (p))
\psi_\downarrow   + \sqrt{\lambda g}(\phi^*\psi_\downarrow \psi_\uparrow +
\phi \psi_\uparrow^\dagger \psi^\dagger_\downarrow) +g \phi^* \phi
\, \, \, \, . \label{2.}
\end{equation}
In the path integral formalism, the generating functional of ($\ref{2.}$) is,
\begin{equation}
Z = \int {\cal D}\psi^\dagger {\cal D} \psi {\cal D} \phi^* {\cal D} \phi
 e^{i\int d^4x {\cal L}_1} \, \, \, \, .
\end{equation}
Performing the fermion functional integral:
\begin{equation}
Z = \int {\cal D}\phi^* {\cal D} \phi e^{iS_{eff} (\phi^*, \phi) }
\, \, \, \, ,
\end{equation}

\noindent
where
\begin{equation}
e^{iS_{\rm eff} [\phi^*, \phi] }   = \int {\cal D}\psi^\dagger
{\cal D} \psi  e^{iS_1} \, \, \, \,
\end{equation}

\noindent
and
\begin{equation}
S_{\rm eff} = -i Tr \log \left( \begin{array} {c c}
p_0 - \epsilon(p) & - \sqrt{\lambda g} \phi \\
\sqrt{\lambda g} \phi^* & p_0 + \epsilon (p) \\
\end{array} \right)
+ \int d^4xg \phi^* \phi  \, \, \, \, .
\end{equation}

\noindent
Here the matrix, if sandwiched between $\psi^\dagger =
(\psi^\dagger_\uparrow, \psi_\downarrow )$ and $\psi = (\psi_\uparrow,
\psi_\downarrow^\dagger )^T$, gives the fermion-dependent part of
($\ref{2.}$). The term without derivatives, is the effective potential for
the condensate and at finite temperature is given by,
\begin{equation}
V_{\rm eff} \sim a \log \frac{T}{T_c} \phi^* \phi + b (\phi^* \phi)^2 \, \, \,
\, ,
\end{equation}
where a, b are positive parameters. It is easily seen that $\phi$ attains a
nonzero VEV below $T_c$, which results in the spontaneous breaking of the
electromagnetic $U(1)$ symmetry and hence superconductivity. In the
following we will proceed along similar lines, keeping in mind the
non-trivial vortex excitations on the plane. \\
The relevant Lagrangian is,
\begin{equation}
{\cal L} = \bar {\psi}_\alpha (i \partial  \! \! \! /
- e A   \! \! \! /) \psi_\alpha - \frac{1}{4\lambda N}
\bar{\psi}_\alpha \psi^c_\alpha \bar {\psi}^c_\beta
\psi_\beta \, \, \, \, .
\end{equation}
Here $\alpha , \beta$ range from $1$ to $N$, $N$ being the parameter
of the large $N$ expansion. Although the coupling constant is
dimensionful, this model is renormalizable in large $N$ perturbation
theory. \\  Introducing the auxiliary fields $\phi$ and $\phi^*$ to
decouple the four fermi term, one gets,

\begin{equation}
{\cal L} = \bar {\psi}_\alpha (i \partial \! \! \! /- e A  \! \! \! /)
\psi_\alpha
+ \frac{1}{2}( \phi^* \bar{\psi}^c_\alpha \psi_\alpha - \phi
\bar{\psi}_\alpha \psi^c_\alpha ) - \lambda N \phi^* \phi \, \, \, \, .
\end{equation}
Proceeding along the same lines as in the BCS case, a derivative expansion
leads to
\begin{equation}
V_{\rm eff} = V^{(0)}  + i N \int \frac{d^3k}{(2 \pi)^3} \log \left(
1 - \frac{\phi^* \phi}{k^2} \right) \, \, \, \, .
\end{equation}

\noindent
The gap equation involving $v_0 \equiv <|\phi|>$,

\begin{equation}
\frac{1}{N} \frac{\delta V_{\rm eff}}{\delta v_0} = 0 =
2 \lambda v_0 - \frac{v_0}{2 \pi}(\sqrt{\Lambda^2 + v^2_0} -v_0) \, \, \, \, ,
\end{equation}
yields a nontrivial VEV for $v_0$ if $\lambda$ is less than a critical value
$\lambda_c \equiv \frac{\Lambda}{4 \pi}$. Notice that the four-fermi coupling
is
$\sim 1/ \lambda$ and hence $\lambda < \lambda_c$ is a strong
coupling regime. The effective potential finally is
\begin{equation}
V_{\rm eff} = N ( \lambda - \lambda_c ) \phi^* \phi + \frac{N}{6\pi}
( \phi^* \phi )^{3/2} \, \, \, \, .
\end{equation}

\noindent
At finite temperature the potential takes the form

\begin{equation}
\frac{1}{N}V_{\rm eff} (\phi, \phi^*;T) = \lambda \phi^* \phi -
\frac{1}{ \beta } \sum_{ -\infty}^\infty \int \frac{d^2 k} {(2 \pi)^2} \log
\left( 1 + \frac{\phi \phi^*} {k^2 + \frac{4\pi^2}{\beta^2}
\left( n+ \frac{1}{2} \right)^2}  \right)  \, \, \, \, .
\end{equation}

\noindent
We analyze the gap equation for a nontrivial constant solution
$v_T \equiv <| \phi |>$:
\begin{equation}
\lambda v_T - \frac{1}{\beta} \sum_{n \epsilon Z} \int \frac{d^2 k}
{(2 \pi)^2} \frac{1}{\left[ {\bf k}^2 + v^2_T + \frac{4\pi^2}{\beta}
\left( n + \frac{1}{2}\right)^2 \right]}  = 0 \, \, \, \, .
\end{equation}

\noindent
Sum over frequencies yields,
\begin{equation}
{v_T}\left(\lambda - \int \frac{d^2k}{(2\pi)^2}
\frac{1}{\sqrt{{\bf k}^2+v_T^2}}
\frac{\sinh(\beta \sqrt{{\bf k}^2+v_T^2})}
{\cosh (\beta\sqrt{{\bf k}^2 + v_T^2)}+1}\right)
= 0 \, \, .
\end{equation}
 From above, defining the critical temperature $T_0$ to be the point
where $v_T = 0$ one gets
\begin{equation}
T_0 = \frac{v_0}{2\log2} \, \, \, \, .
\end{equation}
Hence $T_0$ is the temperature where the neutral order parameter vanishes.
Incorporating electromagnetism, the zero temperature effective Lagrangian
can be computed in a straightforward manner and is given by
\begin{eqnarray}
{\cal L}_{\rm eff} &=& \frac{N}{16\pi v_0}|(\partial_\mu + 2ieA_\mu)\phi|^2
 \cr
&-& \frac{N}{192 \pi v_0^3}(\phi^* \partial_\mu \phi + \phi
\partial_\mu \phi^*) (\phi^* \partial^\mu \phi + \phi \partial^\mu \phi^*)
- V_{\rm eff} \, \, \, \, .
\end{eqnarray}
\noindent
It is worth noting that the effective action does not contain any parity
or time reversal violating terms, e.g.  it can be shown that Chern-Simons
type terms will never be generated. \\
Now the expectation value of a charged order parameter can be easily
computed at finite temperatures. Writing $\phi = \rho e^{i\theta(x)}$
and neglecting fluctuations of $\rho$, one gets
$$
<\phi(x)> = <\rho e^{i\theta(x)}>
\simeq v_T e^{-\frac{1}{2}<\theta^2(x)>}  \, \, \, \, .
$$
$<\theta^2(x)>$ can be computed in a straight forward manner using the
effective action, $^{\rm {4}}$ yielding finally

\begin{equation}
<\phi> ={\rm const}(\beta \eta)^\chi \,  \, \, \, .
\end{equation}

\noindent
Here $\eta$ is an infrared  cutoff and  $\chi = 2/N (\beta v_T \tanh
(\beta v_T/2))$. We see that in the limit $\eta \rightarrow 0, <\phi>
\rightarrow 0$, since $\chi$ is positive. This proves that $\theta$,
the mass less excitation associated with the Goldstone mode, is responsible
for the vanishing of the VEV of the charged order parameter $\phi$ at
nonzero temperatures. \\
Next-to-leading order corrections in the large $N$ perturbation theory shows
that $^{\rm {4}}$ the quantum fluctuations do not destabilize the vacuum
either
due to ultraviolet or infrared effects. \\

\noindent
{\bf 3. ~ ~ Superconductivity due to Vortex binding} \\
We study occurrence of superconductivity in this model using a duality
transformation to take into account the non-trivial topological excitations
on the plane. The phase of the complex order parameter is multivalued and
gives rise to vortex excitations.$^{\rm 10}$ The duality transformation,
allows the effective separation of the single and multivalued components
of the complex field. At low temperature, it will be shown that vortices
and anti-vortices are tightly bound, and the model exhibits
superconductivity. At a critical temperature $T_c$ the celebrated
Kosterlitz-Thouless transition takes place and the vortices become free.
This is precisely the temperature where superconductivity is destroyed. \\
To capture the physics of vortex excitations, we write $\phi = \rho
e^{i \theta}$ with $\rho$ fixed at $v_T$. This is the long wavelength London
limit which neglects fluctuations of $\phi$. Replacing $ \partial_\mu \theta
\rightarrow \partial_\mu \theta - i \varphi^* \partial_\mu \varphi$ with
$\varphi^* \varphi = 1$, we see that $\theta$ describes a single valued field
and $\phi$ accounts for the vortex dynamics. The identically conserved vortex
current is,

\begin{equation}
J^{\rm vort}_{\mu}  = (2 \pi i)^{-1} \epsilon_{\mu \nu \lambda}
\partial^\nu (\varphi^* \partial^\lambda \varphi) \, \, \, \, .
\end{equation}

\noindent
At finite temperature the dynamics is effectively described by$^{\rm 4}$,
\begin{equation}
{\cal L}_{\rm eff} = \frac{N \tanh (\beta v_T/2)}{16 \pi v_T} v^2_T
(\partial^\mu \theta - i \varphi^* \partial^\mu \varphi + 2eA^\mu)^2
\, \, \, \, .
\end{equation}

\noindent
It is useful to rewrite this in terms of the auxiliary current $J_\mu =
\partial^\mu \theta - i \varphi^* \partial^\mu \varphi + 2eA^\mu$:

\begin{equation}
\frac{1}{N} {\cal L}_{\rm eff} = - \frac{1}{2K}J_\mu J^\mu - J_\mu
(\partial^\mu \theta - i \varphi^* \partial^\mu \varphi + 2eA^\mu)
\end{equation}

\noindent
where $ K = \frac{v_T \tanh (\beta v_T/2)}{8 \pi} $. The single valued
field $\theta$ can now be integrated out, giving the constraint,
$\partial^\mu J_\mu = 0$. This is readily solved by
\begin{equation}
J_\mu = \epsilon_{\mu \nu \lambda} \partial^\nu a^\lambda \, \, \, \, .
\end{equation}

\noindent
In terms of this auxiliary gauge field $a_\mu$ the new Lagrangian after
integration by parts becomes,
\begin{equation}
\frac{1}{N} {\cal L}_{\rm eff} = - \frac{1}{4K}f^2_{\mu\nu} -
2 e \epsilon_{\mu \nu \lambda} a^\mu \partial^\nu A^\lambda
-2 \pi a^\mu J^{\rm vort}_\mu \, \, \, \, ,
\end{equation}

\noindent
where $f_{\mu \nu} = \partial_\mu a_\nu - \partial_\nu a_\mu.$ The
integration over $a_\mu$ can now be performed, and the result is

\begin{eqnarray}
{\cal L}_{\rm eff}
= &-& 2 \pi^2 N K J^{\rm vort}_\mu (x) \frac{g^{\mu \nu}}{\partial^2}
J^{\rm vort}_\nu(x) \nonumber \cr
&-& 4 \pi e NK J^{\rm vort}_\mu \frac{g^{\mu \nu}}{\partial^2} \epsilon_{\nu
\lambda \rho} \partial^\lambda A^\rho - e^2 NKF^{\mu \nu} \frac{1}{\partial^2}
F_{\mu \nu} \, \, \, \, .
\end{eqnarray}

\noindent
Neglecting for the moment the second term which describes the interaction
of the electromagnetic field with the vortex, the  current-current
correlation
function is
\begin{equation}
\delta^2 S_{\rm eff} /\delta A_\mu (x) \delta A_\nu (y)
= <j^\mu (x) j^\nu(y)> = -4e^2NK
\left( \frac{\partial^\mu \partial^\nu} {\partial^2} - g^{\mu \nu}\right)
\delta^3(x-y) \, \, \, .
\end{equation}

\noindent
The above expression clearly indicates a pole at zero momentum and hence
superconductivity. In the presence of a Maxwell term, one can show that the
photon propagator, reveals a pole at non-zero momenta which indicates
Meissner effect.

\noindent
As has been seen above, the pole in the current-current correlator arises
because of the $F^{\mu \nu} \partial^{-2} F_{\mu \nu}$ term in the effective
action. At low temperature a slowly varying magnetic field sees pairs of
tightly bound vortices and hence the net contribution from each pair to the
above term vanishes. However above $KT$ phase transition temperature the
vortex contribution to this term {\it exactly cancels the contribution from
the single-valued part,} thereby destroying superconductivity. \\
This can be shown more explicitly, by looking at only the static vortex
configurations ($ J^{\rm vort}_{i} \, \, {\rm and} \, \,
\partial_0 J^{\rm vort}_{0} = 0$) and computing the vortex contribution to
the current-current correlation function. Writing

\begin{equation}
J_0^{\rm vort} \equiv \rho^{\rm vort} (x) =
\sum_a m_a \delta ({\bf x-x}_a(t))
\end{equation}

\noindent
and using the Green's function properties in two dimensions, the effective
action reduces to

\begin{eqnarray}
S_{\rm eff}/N &=& \pi K \beta \sum_{a,a^\prime} m_a m_{a^\prime}
\log |{\bf x}_a -{\bf x}_{a^\prime}| \cr
&+& 2eK \beta \sum_a m_a \int d^2r B(x) \log |{\bf x- x}_a| \, \, \, \, .
\end{eqnarray}

\noindent
Here $m_a$ is the integer valued vorticity and ${\bf x}_a$ is the position
of the $\rm a^{th}$ vortex. Without the interaction term, the above is the
action of the familiar XY model. In the static limit, the contribution of
the vortices to the current-current correlation function is
\begin{eqnarray}
<J^i({\bf q})j^i({\bf -q})>^{\rm vort} &=& \frac{\delta^2 S_{\rm eff}}
{\delta A^i ({\bf q}) \delta A^j ({\bf -q})} \cr
&=& \left( \delta^{ij} - \frac{q^iq^j}{{\bf q^2}}\right) {\bf q^2}
<\rho^{\rm vort} ({\bf q}) \rho^{\rm vort} ({\bf -q})> \, \, \, \, .
\end{eqnarray}

\noindent
In momentum space,
\begin{equation}
\rho^{\rm vort} ({\bf q}) = \sum_a m_a e^{i {\bf q.x}_a} \, \, \, .
\end{equation}

\noindent
and hence in the confined phase,
\begin{equation}
<\rho^{\rm vort} ({\bf q}) \rho^{\rm vort }({\bf -q})> =
\sum_{a,a^\prime} e^{{\bf iq.(x}_a-{\bf x}_{a^\prime})}
<m({\bf x}_a) m({\bf x}_{a^\prime})> \, \, \, \, .
\end{equation}

\noindent
Using the results of the XY model,
\begin{equation}
<m({\bf x}_a) m({\bf x}_{a^\prime})> \sim
\frac{1}{|{\bf x}_a -{\bf x}_{a^\prime}|^{2P}}
\end{equation}
and
\begin{equation}
<\rho^{\rm vort} ({\bf q}) \rho^{\rm vort}({\bf -q})> \sim
\frac{1}{{\bf q}^{2-2P}} ;
\end{equation}

\noindent
where $P = \pi NK \beta$. Since in the confined phase $P >2$, the zero
momentum pole in $<J^i({\bf q})j^i({\bf -q})>$ does not get a contribution
from the confined vortices and hence the Meissner effect is not affected.
In contrast, when the vortex deconfinement, one can directly integrate
out the $J_\mu$ variable and arrive at
\begin{equation}
{\cal L}^{\bf vort}_{\bf eff} = NKe^2 F^{\mu\nu} \frac{1}{\partial^2}
F_{\mu \nu} \, \, \, \, .
\end{equation}

\noindent
This is exactly of opposite sign to that of the contribution of the
single-valued sector. This proves, that in the deconfined phase,
superconductivity is destroyed, and $T_c = T_{KT}$. $T_{KT}$ in this model
can be obtained by solving the following equation
\begin{equation}
\frac{\beta v_T}{2} \tanh \frac{\beta v_T}{2} = \frac{8}{N} \, \, \, \, .
\end{equation}

\noindent
This can be shown to be lower than the temperature where the neutral order
parameter vanishes.

\noindent{\bf 4.~~ Conclusion} \\
To conclude the relativistic four-fermi model presented here, exhibits
superconductivity due to the KT mechanism of vortex-binding. Once the
vortices unbind, the pole in the current-current correlator disappears and
superconductivity is destroyed. This result was anticipated, but not proven,
in some recent works in the literature$.^{\rm 11}$ It would be interesting
to see if in three dimensions, this mechanism generates superconductivity. \\

This paper is dedicated to the memory of Prof.H.Umezawa, who has contributed
significantly to the  understanding of various aspects of superconductivity
and in general to the understanding of the structure of field theories,
both at zero and finite temperatures. \\
\noindent{\bf Acknowledgement}\\
We acknowledge useful discussions with V. Spiridinov, D. Arovas, B. Sakita,
M.B. Paranjape, K. Shizuya and R. Ray. One of the authors (P.K.P.) would like
to thank V. Srinivasan and R.S. Bhalla for many enlightening conversations.
This work was supported in part by the Natural Science and Engineering
Research Council of Canada and the Fonds F.C.A.R. du Qu\'{e}bec.

\noindent
{\bf ~~ References} \\
\begin{enumerate}
\item E.Fradkin, {\it Field Theories of Condensed Matter Systems}
(Addison-Wesley, 1991)
\item  N.D. Mermin and H. Wagner, {\it Phys. Rev. Lett.} {\bf 17, } 1133
(1966); S. Coleman, {\it Commun. Math. Phys. } {\bf 31,} 259 (1973).
\item For references and review, see F. Wilczek, {\it Fractional Statistics
and Anyon Superconductivity} (World Scientific, 1990).
\item  For details of calculation and beyond $1/N$ corrections see,
R. MacKenzie, P.K. Panigrahi and S. Sakhi {\it Phys Rev.} {\bf B48,}
3892 (1993); R. MacKenzie, P.K. Panigrahi and S. Sakhi
{\it Int. J. Mod. Phys.} {\bf A9,} 3603 (1994).
\item  J.M. Kosterlitz and D.J. Thouless, {\it J. Phys.} {\bf C6,} 1181
(1973).
\item  J.B. Marston and Affleck,  {\it Phys. Rev.} {\bf B39,} 11538 (1989).
\item  N. Dorey and N.E. Mavromatos, {\it Nucl. Phys.} {\bf B386,} 614 (1992).
\item  B. Rosenstein, B.J. Warr and S.H. Park, {\it phys. Rep.} {\bf 205,}
59 (1991).
\item B.Sakita, {\it Quantum Theory of Many-Variable Systems and Fields}
(World Scientific, 1985).
\item For a detailed review of the various topological excitations and
other aspects of superconductors both at zero and finite temperatures,
see H. Umezawa, H. Matsumoto, and M. Tachiki, {\it Thermo Field Dynamics}
(North Holland, 1982).
\item A. Kovner and B. Rosenstein, {\it Phys. Rev} {\bf B42} 4748 (1990).
\end{enumerate}

\end{document}